\documentclass[11pt]{article}
\usepackage[psamsfonts]{amssymb}
\begin{document}
\section*{\footnotesize{\it Indian Journal of Theoretical Physics} {\bf 49},
 no. 1, 1-32 (2001)\ \ \ \ \ \ \ \ \ \ \ \ \ \ \ \ \ \ \ \ \ \ \
 \ \ \ \ \ \ \ \ \ \ \ \ \ \ \ \ \ \ \ \ \ \  \ \
 \break (also arXiv.org e-print archive,
           http://arXiv.org/abs/quant-ph/9906091)}

\vspace{6mm}

\bigskip\medskip
\begin{center}
{\LARGE \bf  On the theory of the anomalous photoelectric effect
 stemming from a substructure of matter waves }
\end{center}

\vspace{2mm}
\begin{center}
{\bf Volodymyr Krasnoholovets}
\end{center}

\begin{center}
{Institute of Physics, National Academy of Sciences, \\ Prospect
Nauky 46,   UA-03028 Ky\"{\i}v, Ukraine \\ (web page \
http://inerton.cjb.net)}
\end{center}

\vspace{4mm}
\begin{center}
\ \ \ \ \ \ \ \ \ \ \ \ \ \ \ \ \ \ \ \ \ \ \ \ \ \ \ \ \ \ \ \ \
\ \ \ \ \ \ \ \ \ \ \ \ \ \ \ \ \ \ \ \ \ \ \ \ \ \ \ \ \ \ \ \ \
\ \ \ \ \ \ \ \  February 2000
\end{center}

\begin{center}
{\small{\bf Abstract}}
\end{center}
{\small

The two opposite concepts -- multiphoton and effective photon --
readily describing the photoelectric effect under strong
irradiation in the case that the energy of the incident light is
essentially smaller than the ionization potential of gas atoms and
the work function of the metal are treated. Based on the
submicroscopic construction of quantum mechanics developed in the
previous papers by the author the analysis of the reasons of the
two concepts discrepancies is led. Taking into account the main
hypothesis of those works, i.e., that the electron is an extended
object that is not point-like, the study of the interaction
between the electron and a photon flux is carried out in detail. A
comparison with numerous experiments is performed. }

\vspace{3mm}

\begin{description}
\item [{\bf Key words:}] space, matter waves, inertons,
laser radiation, photoelectric  effect       \\ \vspace{2mm}

\item[\sf {\bf PACS:}]
03.75.-b Matter waves. \ 32.80.Fb Photoionization
 of atoms and ions. \ \ \ \ \ \ \ \break 42.50.Ct Quantum description
 of interaction of light and matter; related experiments

\end{description}

\newpage
\section{\bf Introduction}

\hspace*{\parindent}
   The previous papers of the author [1-3] present a quantum theory
operating at the scale $\sim 10^{-28}$ cm (this size combines all
types of interactions as required by the grand unification of
interaction). The theory takes into account such general
directions as: deterministic view on quantum mechanics pioneered
by L.  de Broglie and D. Bohm (see, e.g. Refs. 4,5), the search
for a physical vacuum model in the form of a real substance (see,
e.g. Refs. 6-11), the introduction in the united models as if a
"superparticle" whose different states are the electron, muon,
quark, etc. [12], and the model of polaron in a solid, i.e., that
a moving charged carrier strongly interacts with a polar medium.

    The kinetics of a particle constructed in works [1-3] easily
results in the Schr\"odinger and Dirac formalisms at the atom
scale. Besides the developed theory could overcome the two main
conceptual difficulties of standard nonrelativistic quantum
theory. First, the theory advanced a mechanism which could
naturally remove long-range action from the nonrelativistic
quantum mechanics. Second, the Schr\"odinger equation gained in
works [1,2] is Lorentz invariant owing to the invariant time
entered the equation.

     The main distinctive property of the theory was the prediction of
special elementary excitations of space surrounding a moving
particle. It was shown that space should always exhibit resistance
to any canonical particle when it starts to move: the moving
particle rubs against  space and such a friction generates virtual
excitations called "inertons" in papers [1,2].  Thus the inerton
cloud around a moving particle one can identify with a volume of
space $\mathfrak{V}$ that the canonical particle excites at its
motion. In other words, the inerton cloud may be considered as a
substructure of the matter waves which are described by the wave
$\psi$-function in the region of the $\mathfrak{V}$.

     Nonetheless, the question arises whether one can reveal a cloud
of inertons, which accompany a single canonical particle. As was
deduced in Ref. 1, the amplitude of spatial oscillations of the
inerton cloud $\Lambda /\pi$ correlates with the amplitude of
spatial oscillations of the particle, that is, the de Broglie
wavelength of the particle $\lambda$:
\begin{equation}
         \Lambda = \lambda {\kern 1pt}c/v_0
\label{1}
\end{equation}
where $v_0$ is the initial velocity of the particle and $c$ is the initial
velocity of inertons (velocity of light). If $v_0 \ll c$ then
$\Lambda \gg \lambda$ and hence the disturbance of space in the form
of the  inerton cloud should appear in an extensive region around the
particle.  In this connection, the cloud of inertons may be detected,
for instance, by applying a high-intensity luminous flux.

To examine this assertion, let us turn to experimental and
theoretical results available when laser-induced gas ionization
phenomena and photoemission from a laser-irradiated metal take
place.

\section {\bf The two opposite concepts}

\hspace*{\parindent}
    First reports on the experimental demonstration of the
laser-induced gas ionization occurrence at a frequency below the
threshold appeared in the mid-1960s (Meyerand and Haught [13],
Voronov and Delone [14], Smith and Haught [15] and others). Those
works launched detailed experimental and theoretical study of the
new unexpected phenomena. At the time being, it would seem the
mechanism providing the framework for the phenomena has been
roughly understood. However, this is not the case: all materials
available one can subdivide into two different classes.

    Having taken a critical view of the effect in which the photon
energy of the incident light is essentially smaller than the
ionization potential of atoms of rarefied noble gases and the work
function of the metal, we shall turn to the two opposite
standpoints excellently expounded in the reviews by Agostini and
Petite [16] and Panarella [17,18]. At the same time it should be
particularly emphasized that any improvement of the multiphoton
theory is not the aim of the present work. The author wishes only
to show that something more fundamental is hidden behind the
formalism of orthodox quantum mechanics that is employed as a base
for the study of a matter irradiated by an intensive light.

\subsection {\it Multiphoton concept}

\hspace*{\parindent}
 The review paper by Agostini and Petite [16]
analysed several tens of works exploiting the prevailing
multiphoton theory. The multiphoton concept is based on the
typical interaction Hamiltonian
\begin{equation}
{\hat H}_{\rm int}= -e{\hat {\vec z}}{\vec E}_0 \cos \omega t
\label{2}
\end{equation}
which specifies the interaction between the dipole moment $e \vec
z$ of an atom and the incident electromagnetic field $\vec E
={\vec E}_0 \cos \omega t$. The concept starts from the standard
time dependent perturbation theory, Fermi [19], describing a
probability per unit time of a transition of an atom from the
bound state $|i>$  to a state $<{\rm c}|$ in the continuum. On the
next stage the concept modifies the simple photoelectric effect to
the nonlinear one (see, e.g.  Keldysh [20] and Reiss [21]) in
which the atom is ionized by absorption of several photons. The
$\cal N$th-order time dependent perturbation theory changes the
usual Fermi golden rule to $\cal N$ photon absorption that
produces the probability [16]
\begin{equation}
w_{\cal N} = \frac{2\pi}{\hbar}\Bigl( \frac {2e^2}{\varepsilon_0 c}
\Bigr)^{\cal N} \sum_{\rm c} \Big| \sum_{i, j, ..., k} \frac
{<g|z|i><i|z|j>... <k|z|{\rm c}>}{(E_g + \hbar \omega - E_i)(E_g
+2\hbar\omega - E_j)...} \Big|^2
\label{3}
\end{equation}
where $|i>, \ |j>, \ ... ,\ |k>$ are the atomic states, $I$ the
intensity of laser beam and $|{\rm c}>$ the continuum states with
energy $E_g + \hbar \omega, \ E_g $ being the energy of the ground
state $|g>$. The summation over intermediate states could be
performed by several methods. An estimation of the probabilities
of multiphoton processes can be made utilizing the so-called
generalized cross section [16]
\begin{equation}
s_{\cal N}= 2\pi (8\pi \alpha)^{\cal N} r^{\kern 1pt 2 {\cal
N}}{\kern 1pt}\omega^{-{\cal N}+1} \label{4}
\end{equation}
where $r \sim 0.1$ nm is the effective atom radius and $\alpha
=1/137$ is the fine structure constant. The Einstein law $E=\hbar
\omega$ characterizing the simple photoelectric effect changes to
the relation specifying the nonlinear photoelectric effect
\begin{equation}
E_{\rm c}={\cal N} \hbar \omega - E_i. \label{5}
\end{equation}

    The $\cal N$-photon ionization rate (3) is proportional to
$I^{\cal N}$. This prediction, as was pointed out by Agostini and
Petite [16], verified experimentally up to ${\cal N}=22$ and with
laser intensity up to $10^{15}$ W/cm$^2$. They noted that "$I$
must be below the saturation intensity to perform this
measurement. When $I$ approaches to $I_{\rm s}$, one must make out
account the depletion of the neutral atom population, which
modifies the intensity dependence of the ion number". It may be
seen from the preceding that $I_{\rm s}\geq 10^{15}$ W/cm$^2$.

      At the same time we should note that the experiment does not
point clearly to the dependence of $I^{\cal N}$. The experiment
only demonstrates that in a log-log plot $N_i$ versus light
intensity $I$ where $N_i$ is the number of ionized atoms of gas
all points are located along a straight line whose slope is
proportional to $\cal N$. This was shown by Lompre {\it et al.}
[22] for Xe, Kr, and Ar with an accuracy about 2 $\%$ . Such
result was interpreted [22] as a simultaneous absorption of $\cal
N$ photons; the linear slope was held to $2 \times 10^{13}$
W/cm$^2$ and the maximum value was ${\cal N} = 14$.

     The authors of the review [16] marked that good agreement the
multiphoton theory and experiment had till the experimental
investigation (Martin and Mandel [23] and Boreham and Hora [24])
of the energy spectra of electrons ejected in the ionization of
atoms; the kinetics energy of ejected electrons was far in excess
of the prediction. Since then the multiphoton concept has advanced
to so-called the above-threshold ionization (ATI). It replaced
relationship (5) for
\begin{equation}
E_{\rm c}=({\cal N}+S)\hbar \omega - E_i
\label{6}
\end{equation}
where $S$ is the positive integer. Several consequences were
checked experimentally: branching ratios (Petite {\it et al.} [25]
and Kruit {\it at el.} [26]), the intensity dependence, i.e.,
proportionality to $I^{{\cal N}+S}$ (Fabre {\it et al.} [27],
Agostini {\it et al.} [28] and others [16]).

     An attempt to verify the nonlinear photoelectric effect on metals
was undertaken by Farkas [29] (however, see below).

     During the last decade a number of further studies of the
multiphoton ionization of atoms under ultra intensive laser
radiation have been performed both experimentally and
theoretically (see, e.g. review papers and monographs [30-37]).
For example, papers of Avetissian {\it et al.} [35,38] deal with
the relativistic theory of ATI of hydrogen-like atoms; at the same
time the authors note that the idea of introducing of the
stimulated bremsstrahlung for the description of the photoelectron
final state still remains as a great problem for the ATI process.
Besides the definition of wave dynamic function of an ejected
electron stands problematic as well.

Unfortunately the major deficiency of the ATI and more advanced
models is too complicated expressions for the probability of
ejected photoelectrons. Such expressions need additional
assumptions. Hence a distinguish feature of the nonlinear
multiphoton theory is the availability of a great many free
parameters. Besides all the recent experiments operate with
extremely short laser pulses which rather strikes atoms then
slowly excite them. And this has cast some suspicion on the
application of the time dependent perturbation theory
(nonrelativistic or relativistic) for the description of ejection
of photoelectrons from atoms in all cases. More likely femtosecond
laser pulses create new effects which need new detailed studies
(such as the scattering of electrons radiated from atoms
immediately after ionization that tries to account the eikonal
approximation [38], etc.).

    Thus the results obtained with different lasers might be
different as well. Below we will analyze only the pure multiphoton
concept that became the starting point for the further
complications; in other words we will treat the case of the
adiabatic turning on electromagnetic perturbation. Notwithstanding
the fact that the multiphoton methodology is widely recognize
today, we should emphasize that it ignored some "subtle"
experimental results obtained with the use of nanosecond and
picosecond light/laser pulses in the 1960s and 1970s (perhaps
setting that such results were caused by indirect reasons).

\subsection {\it Effective photon concept}

\hspace*{\parindent} In review papers Panarella [17,18] analysed
about a hundred of other experiments devoted a laser-induced gas
ionization and laser-irradiated metal. Panarella explicitly
described all dramatic events connected with the construction of a
reasonable mechanism which could explain unusual experimental data
on the basis of standard concepts of quantum theory. Based on
those experimental results he convincingly demonstrated the
inconsistency of generally accepted multiphoton methodology. In
particular, Panarella studied the following series of experiments:
1) variation of the total number  $N_i$ of ionized gas atoms as a
function of the laser intensity $I_{\rm p}$ (see Refs. 17,18 and
also Agostini {\it et al.} [39]). In a log-log plot the
experimental points did not lie on a straight line and the
inflection point, for all gases studied, got into the range
approximately from $10^{12}$ to $10^{13}$ W cm$^{-2}$ at the laser
wavelength 1.06 $\mu$m and from $10^{11}$ to $10^{12}$ W/cm$^2$ at
the laser wavelength 0.53 $\mu$m (note that such an inflection
point, as was mentioned in the previous subsection, should refer
to the saturation intensity whose value $I_{\rm s}$, however, of
the order of $10^{15}$ W/cm$^2$!); 2) variation of the total
number $N_i$ as a function of time $t$ of the increase in
intensity of laser pulse (the experiment by Chalmeton and Papoular
[40]); 3) variation of the breakdown intensity threshold against
the gas density (see  experiments by Okuda {\it et al.} [41-43]);
4) focal volume dependence of the breakdown threshold intensity
(see, e.g. the experiment by Smith and Haught [15]); and others.

    All those experiments could not be explained in the framework
of the multiphoton methodology. The multiphoton concept failed to
interpret just fine details revealed in the experiments. Among
other things Panarella stressed that the experiment by Chalmeton
and Papoular [40] was a crucial one.

  The cascade theory (see, e.g. Zel'dovich and Raizer [44])
was also untenable to explain a number of data (see Ref. 17). This
theory conjectured that random free electrons with the great
energy were present in the gas and those electrons along with
newly formed electrons generated other electrons; it was conceived
that the optical field accelerated the electrons.

    Panarella analysed several other theoretical hypotheses which
assumed the existence of high-than-normal energy photons in laser
beam: the model based on quantum formalism, Allen [45], the model
based on quantum potential theory, Dewdney {\it et al.} [46,47],
and the model resting on classical electromagnetic wave theory of
laser line broadening, de Brito and Jobs [48] and de Brito [49].
The first two models operated with the Heisenberg uncertainty
principle and de Broglie-Bohm quantum potential respectively; it
was expected that the deficient energy of a photon could appear
due to some quantum effects. The last model suggested that the
existence of separate high-energy photons in the laser beam might
be stipulated by the laser line shape. Unfortunately the models
could not explain the whole series of available experimental
results.

    In contrast to those concepts, Panarella noted [17] that new physics
should be present in the phenomena described above and proposed an
effective photon theory [17,18]. He postulated that the photon
energy expression $\varepsilon =h\nu$ had to be modified "ad hoc"
into the novel one:
\begin{equation}
\varepsilon = \frac {h\nu}{1-\beta_{\nu}f(I)}
\label{7}
\end{equation}
where $f(I)$ is the function of the light intensity and
$\beta_{\nu}$ is a coefficient. In this manner Panarella's theory
holds that, at the extremely high intensities of light,
photon-photon interaction begins to play a significant role in the
light beam such that the photon energy becomes a function of the
photon flux intensity. To develop an effective photon concept it
was pointed out [18] that the number density of photons in the
focal volume is much larger than ${\widetilde \lambda}^{-3}$ where
$\widetilde \lambda$ is the wavelength of laser's irradiated
light. In this respect he came up with the proposal to reduce the
photon wavelength in the focal volume. He assumed that it
unquestionably followed from quantum electrodynamics that photons
could not come any closer than $\widetilde \lambda$.

The effective photon concept satisfied all available experimental
facts mentioned above in this subsection. Moreover the concept was
successfully applied to Panarella's own first-class experiments on
electron emission from a laser irradiated metal surface [50,51,18]
and to other experiments (Pheps [52] and see also Refs. 17,18).

   Such remarkable success of the formula (7) gave rise to the confidence
that some hidden reasons could be a building block for
understanding the principles of effective photons formation [18].
An elementary consideration of photons and hence effective photons
based on neutrinos has been constructed by Raychaudhuri [53].

\vspace{4mm}

Thus in this section we have given an objective account of facts
and adduced the two absolutely opposite views on the same
phenomena. So we need to establish the reasons for the main
discrepancies between the multiphoton and effective photon
concepts and then develop an approach that would reconcile them.

\section{\bf Interaction between the photon flux and an
          electron's inerton cloud}

 \hspace*{\parindent}
First of all we need to discuss in short such notions as the
photon and photon flux.  On question, what is photon?, quantum
electrodynamics answers (see, e.g. Berestetskii {\it et al.}
[54]): it is something that can be described by the equation
\begin{equation} \partial^2 \vec
A /\partial t^2 - c^{-2} \partial^2\vec A/\partial {\vec r}^2=0
\label{8}
\end{equation}
where $\vec A$ is the vector potential that satisfies the condition
\begin{equation}
{\rm div}  {\vec A}=0.
\label{9}
\end{equation}

    The vector potential operator $\hat{\vec A}$
of the free electromagnetic field is constructed in such a way
that each wave with a wavevector $\vec q$ corresponds to one
photon with the energy $h\nu_{\vec q}$ in the volume $V$, that is,
$\hat{\vec A}$ is normalized to $V$ in accordance with the formula
(see, e.g. Davydov [55,56])
\begin{equation}
{\hat {\vec A}}(\vec r, t) = \sum_{\vec q, \alpha} \Bigl( \frac
{ch}{|\vec q| V}\Bigr)^{1/2} \ e^{i\vec q \vec r}{\kern 2pt} {\vec
j}_{\alpha} \ (\vec q) \ \Bigl({\hat a}_{\vec q \alpha}(t) + {\hat
a}^{\dagger}_{-\vec q \alpha} (t) \Bigr) \label{10}
\end{equation}
where $c$ is the velocity of light, $h$ is Planck's constant,
$\vec q$ is the wave vector ($|\vec q| = 2 \pi/{\widetilde
\lambda}$), ${\vec j}_{\alpha} (\vec q)$ is the unit vector of the
$\alpha$th polarization, ${\hat a}^{\dagger}_{\vec q \alpha}$
(${\hat a}_{\vec q \alpha}$) is the Bose operator of creation
(annihilation) of a photon, and $V$ is the volume containing the
electromagnetic field.

A pure particle formalism can also be applied to the description
of the free electromagnetic field; in this case each of the
particles -- photons -- has the energy $\varepsilon =h\nu$ and the
momentum $\hbar{\vec k}(=h\nu/c)$. Just such "photon language" is
often more convenient. It admits to consider a monochromatic
electromagnetic field as a single mode which contains a number of
photons.

    Now let us start by considering the origin of the disagreements
between the two opposite concepts.

That was considerable success of the multiphoton concept that it
incorporated $\cal N$ photons whose total energy was equal to the
potential of ionization of an atom, expression (5). A prerequisite
for the construction of the concept was the supposition that there
was strong nonlinear interaction between a laser beam and a gas.

{\it Criticism}: The multiphoton methodology does not take into
account the threshold light intensity needed for gas ionization.
The photoelectric effect, as such, is not investigated, the
methodology only suggests that atoms of gas may be excited to the
energy level (5) in the continuum. Besides the methodology ignores
the fact of the coherence of the electromagnetic field irradiated
by laser. At the same time the problem of electromagnetic
radiation may be reduced to the problem of totality of harmonic
oscillators, ter Haar [57], which in the case of the laser
radiation must be regarded as coherent. This means that each of
the $\cal N$ photons absorbed should have the same right, but
using the $\cal N$th-order time dependent perturbation theory one
adds photons successively. (The distinction between the incoherent
and coherent electromagnetic field is akin to that between the
normal and superconducting state of the same metal in some sense.
Indeed in a superconductor electrons can not be considered
separately: all superconducting phenomena are caused by cooperate
quantum properties of electrons. That is why describing
superconducting phenomena one should include the cooperation of
electrons, for instance the Meissner-Ochsenfeld effect.)

The advantages of the effective photon concept are its flexibility
at the analysis of experimental results. The concept assumed the
existence of the threshold light intensity that launches
ionization of atoms of gas and ejection of electrons from the
metal. The effective photon was deduced from the assumption that
there could not more than one orthodox photon in a volume of space
$\sim {\widetilde \lambda}^3$. Owing to the huge photon density in
the laser pulse the concept conjectured that photons could
interact with each other forming "effective photons" (7). The
latter are absorbed as the whole and the absorption is a linear
process, which is highly similar to the simple photoelectric
effect.

{\it Criticism}: Photons are subjected to Bose-Einstein statistics
and this means that it is not impossible that the volume $V$
contains an enormous number of photons with the same energy $h
\nu_{\vec q_0}$. In other words, the density of photons depends on
the initial conditions of the electromagnetic field generation. In
any event the statistics is absolutely true at the atom (and even
nucleus) scale, i.e., so long as the photon concentration in the
pulse does not far exceed the concentration of atoms in a solid
$\sim 10^{23}$ cm$^{-3}$. (Note that a somewhat similar pattern is
observed  when  the intensity of sound in a crystal is enhanced.
In the original state acoustic phonons obey the Planck
distribution, but when  the ultrasound is switched on, the phonon
density increases while the volume of the crystal remains the
same.)

Having described ionization of atoms of gas and photoemission from
a metal in terms of the submicroscopic approach [1-3], an effort
can be made to try to develop a theory of the anomalous
photoelectric effect in which electron's wide spread inerton cloud
simultaneously absorbs a number of coherent photons from the
intensive laser pulse.  Thus the theory will combine Panarella's
idea on the anomalous photoelectric effect and the idea of the
multiphoton concept on simultaneous absorption of $\cal N$
photons.

We shall assume that in the first approximation atoms of gas and
the metal may be considered as systems of quasi-free electrons.
The Fermi velocity of $s$ and $p$ electrons in an atom is equal to
(1-2)$\times 10^8$ cm/s. Setting $v_{\rm F}=v_0 \simeq 2\times
10^8$ cm/s one obtains $\lambda =h/mv_0 \simeq 0.36$ nm ($m$ is
the electron mass) and then in accordance with relation (1) the
amplitude of oscillations of the inerton cloud equals $\Lambda
/\pi \simeq 17$ nm. The cloud has anisotropic properties: it is
extended on $\lambda $ along the electron path, that is, along the
velocity vector $\vec v_0$, and on $2\Lambda/\pi$ in the
transversal directions. This means that the cross section $\sigma$
of the electron together with its inerton cloud in the systems
under consideration should satisfy the inequalities:
\begin{equation} \frac{\lambda^2}{4\pi}< \sigma <
\frac{\Lambda^2}{\pi}, \ \ \ \ \ \ \ {\rm or } \ \ \ \ 10^{-16} \
{\rm cm}^2< \sigma < 1.7 \times 10^{-12} \ {\rm cm}^2; \label{11}
\end{equation}
here one takes into account that the radius of electron's inerton
cloud equals $\Lambda/\pi$. At the same time the cross-section of
an atom is only $\sim 10^{-16}$ cm$^2$. The intensity of light in
(10-100)-psec focused laser pulses used for the study of gas
ionization and photoemission from metals was of the order of
$10^{12} - 10^{15}$ \ W/cm$^2$, that is, $10^{30} - 10^{33}$
photons/cm$^2$ per second. Dividing this intensity into the
velocity of light one obtains the concentration of photons in the
focal volume $n\simeq 3\times (10^{19} - 10^{22})$ cm$^{-3}$ and
hence the mean distance between photons is $n^{-1/3}\simeq (30 -
3)$ nm. The number of photons bombarding the inerton cloud around
an individual electron is $\sigma n^{2/3}$; this value can be
estimated, in view of inequality (11), as
\begin{eqnarray}
&1&< \sigma n^{2/3} < 10^3 \ \ \ {\rm at} \ \ \  n \approx 3\times
10^{19} \ {\rm cm}^{-3} \ \ \ \ \  {\rm and}     \nonumber   \\
&1&< \sigma n^{2/3}< 10^5 \ \ \ {\rm at} \ \ \ n \approx 3 \times
10^{22} \ {\rm cm}^{-3}.  \label{12} \end{eqnarray}

     The next thing to do is to write the model interaction between the
electron inerton cloud and an incident coherent light. In an
ordinary classical representation the electron in the applied
electromagnetic field is characterized by the energy
\begin{equation}
{\cal E}= \frac 1{2m}(\vec p - e \vec A)^2
\label{13}
\end{equation}
where $\vec A$ is the vector potential of the electromagnetic
field. This usually implies that the vector potential $\vec A$ in
Amp\'ere's formula (13) relates to the field of one photon. This
is confirmed by expression (11) and the supposition that the
electron can be considered as a point in its classical trajectory.
In the language of quantum theory this means that both the wave
function of the electron and the wave function of the photon are
normalized to one particle in the same volume $V$, Berestetskii
{\it et al.} [58]. However, as follows from the analysis above,
the electron jointly with its inerton cloud is an extended object.
Because of this, it can interact with many photons simultaneously
and the coupling function between  the electron and the applied
coherent electromagnetic field should be defined by the density of
the photon flux. Therefore, contrary to the usual practice to use
the approximation of single electron-photon coupling  (13) in all
cases, one can introduce the approximation of the strong
electron-photon coupling
\begin{equation} {\cal E}= \frac 1{2m} (\vec p
- e {\vec A}_{\rm eff})^2
\label{14}
\end{equation}
which should be correct in the case of simultaneous
absorption/scattering of $\cal N$ photons  by the electron.
Thus in (13)
\begin{equation} {\vec A}_{\rm eff} =e \vec A {\cal N}, \ \ \ \ \ \ \
{\cal N}=\sigma n^{2/3}.
\label{15}
\end{equation}

      In experiments involving noble gases discussed by
Panarella [17,18] the laser pulse intensity had the triangular
shape. We shall apply the same approach. In other words, let the
intensity be changed over the duration $\Delta t$ of the laser
pulse whose intensity runs along the two equal sides of the
isosceles triangle, that is from $I=0$ at $t=0$ to the peak
intensity $I=I_{\rm p}$ at $t=t_{\rm p}=\Delta t/2$ and then to
$I=0$ at $t=\Delta t$.

    Thus ${\vec A}_{\rm eff}$ becomes time dependent;
it can be present in the form
\begin{equation}
{\vec A} _{\rm eff}(\vec r, t) = {\vec A}_{\rm p}{\kern 1pt} e^{i
\vec k \vec r - i\omega t} {\cal N}(t) \label{16}
\end{equation}
where ${\vec A}_{\rm p}$ is the vector potential of the electromagnetic
field at the peak intensity of the pulse,
\begin{equation}
{\cal N}(t)= \sigma_{\rm th} n^{2/3}_{\rm th}\frac{t}{t_{\rm p}}
\label{17}
\end{equation}
is the number of photons absorbed by the electron where
$n^{2/3}_{\rm th}$ is the effective photon density in the unit
area at the threshold intensity of the laser pulse when the energy
of $\sigma_{\rm th} n^{2/3}_{\rm th}$ photons reaches the absolute
value of the ionization potential of atoms or the work function of
the metal, that is $h\nu \sigma_{\rm th} n^{2/3}_{\rm th}= W$. As
relation (17) indicates the cross section of electron's inerton
cloud, $\sigma$ is also signed by dependence on the threshold
intensity; much probably $\sigma$ is not constant and depends on
the velocity of the electron, the frequency of incident light and
the light intensity. The presentation (16) is correct within the
time interval $\Delta t/2$, that is, $t \in [0, t_{\rm p}]$.

     Hence passing on to the Hamiltonian operator of the electron in the
intensive field one has
\begin{equation}
\hat H =\frac{\hat {\vec p}^{\ 2}}{2m} - \frac{e}{m}{\vec A}_{\rm eff}(r,t)
\ {\hat {\vec p}};
\label{18}
\end{equation}
here we are restricted to the linear field effect, much as it is
made in the theory of simple photoelectric effect (see, e.g.
Berestetskii {\it et al.} [58], Blokhintsev [59], and Davydov
[60]).

      In the case of the simple photoelectric effect the Sch\"odinger
equation for the electron
\begin{equation}
i\hbar \frac{\partial \psi}{\partial t}=({\hat {\cal H}}+ {\hat {\cal W}}
(\vec r,t) ) \psi
\label{19}
\end{equation}
contains the Hamiltonian operator ${\hat {\cal H}}$ of the electron in an
atom (or the metal) and the interaction operator
\begin{equation}
{\hat {\cal W}}(\vec r, t) =-\frac{e}{m}\vec A (\vec r, t) \ {\hat {\vec p}}
\label{20}
\end{equation}
whose matrix elements are much smaller than those of the operator
$\hat{\cal H}$. However in our case the matrix elements of the operator
\begin{equation}
{\hat {\cal W}}_{\rm eff}(\vec r, t)= - \frac {e}{m}{\vec A}_{\rm eff}
(\vec r, t) \ {\hat {\vec p}}
\label{21}
\end{equation}
do not seem to be small due to the great value of ${\vec A}_{\rm p}$.
Therefore, exploiting the perturbation theory, we should resort to
the procedure, which makes it feasible to extract a small parameter.

       Nonetheless, the necessary smallness is already inserted into the
structure of the vector potential ${\vec A}_{\rm eff}(\vec r, t)$: the
number of photons absorbed by the electron is a linear function of the
duration of the growing intensity of the pulse [see (17)].
Consequently the interaction operator (21) can be safely used
for $t\ll t_{\rm p}$.

\vspace{4 mm}

\section{\bf Anomalous photoelectric effect}

 \hspace*{\parindent}
   In the absence of the external field the Schr\"odinger equation
\begin{equation}
i\hbar \frac {\partial \psi_0}{\partial t}={ \hat{\cal H}} \psi_0
\label{22}
\end{equation}
which describes the electron (in an atom or metal) has the solution
\begin{equation}
\psi_0 = e^{-i \frac{{\hat{\cal H}}}{\hbar}t}.
\label{23}
\end{equation}
Eq. (22) is transformed in the presence of the field to the equation
\begin{equation}
i\hbar \frac {\partial \psi}{\partial t}= ({\hat {\cal H}}+ {\hat {\cal W}}
_{\rm eff}(t))\psi.
\label{24}
\end{equation}
The $\psi$ function  from (24) can be represented in the form
(see, e.g. Fermi [19])
\begin{equation}
\psi (\vec r, t) =e^{- \frac {\hat {\cal H}}{\hbar} t}\sum_l
a_l (t) \psi_l (\vec r);
\label{25}
\end{equation}
here $a_l (t)$ are coefficients at eigenfunctions $\psi_l (\vec
r)$. By substituting function (25) into Eq. (24)  and multiplying
a new equation by $\psi^*_f (\vec r)$ to left and then integrating
over $\vec r$ one obtains
\begin{equation}
i\hbar \frac { \partial a_f (t)}{\partial t} = \sum_l a_l(t)<f|{\hat
{\cal W}}_{\rm eff}|l> e^{i \omega_{fl}t}
\label{26}
\end{equation}
where $\hbar \omega_{fl}= E_f - E_l$, $E_{f(l)}$ is the eigenvalue of
Eq.  (22) and the matrix element
\begin{equation} <f|{\hat {\cal
W}}_{\rm eff}|l>  = - \frac{e}{m}\int \psi^*_f \ {\vec A}_ {\rm
eff}(\vec r, t) \ {\hat {\vec p}} \ \psi_l \ d \vec r.
\label{27}
\end{equation}
In the first approximation  the coefficient is equal
\begin{equation} a^{(1)}_{f} \simeq \frac 1{i \hbar} \int\limits^t_0
<f| {\hat {\cal W}}_ {\rm eff}| l> e^{i \omega_{fl} \tau} d \tau.
\label{28}
\end{equation}
The possibility of the transition from the atomic state $E_l$ to
the state of ionized atom $E_f$ (or the possibility of the
ejection of electron out of the metal) is given by the expression
\begin{equation}
P(t) \equiv P_f(t)= |a^{(1)}_f(t)|^2,
\label{29}
\end{equation}
or in the explicit form
\begin{equation}
P(t) = \Big| \frac 1{i\hbar} \frac {e}{m}<f|{\vec A}_{\rm p}
{\hat {\vec p}}|l>  \Big|^2 \ \
\Big|  \int\limits^t_0{\cal N}(\tau) e^{i(\omega_{fl} - \omega) \tau}
d \tau    \Big|^2.
\label{30}
\end{equation}

         The first factor in (30) is well known in the simple
photoelectric effect, because it defines the probability of the
electron transition from the atomic $|l>$ to the free state $<f|$.
This factor can be designated as $|M|^2$ and extracted from (30)
in the explicit form (see, e.g. Blokhintsev [59]):
\begin{eqnarray}
|M|^2 \equiv  \Big| \frac 1{i \hbar}<f|{\hat {\vec A}}_{\rm p}
{\hat {\vec p}} |l>  \Big|^2
\ \ \ \ \ \ \ \ \ \ \ \ \ \ \ \ \ \ \ \ \ \ \ \ \
\nonumber  \\
 =16 \pi \ \frac{e^2 \hbar^2}{m^2 V} \ {\vec A}^{\kern 1.5pt 2}_{\rm p} \
\bigl( \frac Z{a_{\rm Bohr}} \bigr)^5  \ \bigl( \frac \hbar {\vec
p}_{\rm free} \bigr)^6 \ \frac{\sin^2 \theta \cos^2 \phi}{(1 -
\frac{v}{c}\cos \theta)^4};
\label{31}
\end{eqnarray}
here $V$ is a normalizing volume, $a_{\rm Bohr}$ is Bohr's radius,
$Z$ is the number charge, ${\vec p}_{\rm free}$ is the momentum of
the stripped electron. The last factor in (31) shows that the
momentum ${\vec p}_{\rm free}$ falls within the solid angle $d
\Omega$ ($v$ is the velocity of the free electron and $|{\vec
p}_{\rm free}|=mv$). Taking into account that the vector potential
$\vec A$ of the electromagnetic field is connected with the
intensity $I$ of the field through the formulas
\begin{equation}
I=\varepsilon_0 c^2 |\vec E|^2, \ \ \ \ \ \ \ \  \vec
E =- \frac{\partial \vec A}{\partial t} = i \omega {\vec A}_{\rm p}
e^{i (\omega t - \vec k \vec r)} \label{32}
\end{equation}
we gain the relation
\begin{equation} {\vec A}^{\kern 1.5pt 2}_{\rm p} = \frac 1{\varepsilon_0
c^2 \omega^2 } \ I_{\rm p}.  \label{33}
\end{equation}
The intensity $I_{\rm p}$ can be separated out of the matrix
element (31), i.e., we can write
\begin{equation} |M|^2 = |{\cal M}|^2 \ I_{\rm p}
\label{34}
\end{equation}
where
\begin{equation}
|{\cal M}|^2 =16 \pi \ \frac {e^2 \hbar^2} {\varepsilon_0 c^2
\omega^2 m^2 V} \ {\vec A}^{\kern 1.5pt 2}_{\rm p} \ \bigl( \frac
Z{a_{\rm Bohr}} \bigr)^5  \ \bigl( \frac \hbar {\vec p}_{\rm free}
\bigr)^6 \ \frac{\sin^2 \theta \cos^2 \phi}{(1 - \frac{v}{c}\cos
\theta)^4}; \label{35}
\end{equation}

   Now, expression (30) can be rewritten as
\begin{equation}
P(t)=|{\cal M}|^2 \ I_{\rm p} \   |{\cal I}(t)|^2
\label{36}
\end{equation}
where
\begin{equation}
|{\cal I}(t)|^2 = \int\limits^t_0 {\cal N}^* (\tau) e^{-i (\omega_{fl}
-\omega)\tau } d \tau
 \int\limits^t_0 {\cal N}(\tau) e^{i(\omega_{fl} -\omega)\tau } d \tau.
\label{37}
\end{equation}
Let us calculate the integral ${\cal I}(t)$:
\begin{eqnarray}
{\cal I}(t) =  \int\limits^t_0 {\cal N} (\tau) e^{i(\omega_{fl}
-\omega)\tau} d \tau = \frac {\sigma_{\rm th} n^{2/3}_{\rm
th}}{t_{\rm p}} \int\limits^t_0 \tau e^{i(\omega_{fl}- \omega)
\tau} d \tau \ \ \ \ \ \ \ \ \ \ \ \ \ \ \ \      \nonumber  \\ =
\frac{\sigma_{\rm th} n^{2/3}_{\rm th}}{t_{\rm p}} {\kern
1pt}[\frac {t}{i(\omega_{fl} - \omega)} {\kern
1pt}e^{i(\omega_{fl} - \omega)t} + \frac 1{(\omega_{fl}
-\omega)^2} {\kern 1pt} (e^{i(\omega_{fl} - \omega)t} - 1) ]. \
\label{38}
\end{eqnarray}
Substituting ${\cal I}(t)$ and ${\cal
I}^*(t)$ into (37) one obtains
\begin{eqnarray} |{\cal I}(t)|^2 =
\frac{(\sigma_{\rm th} n^{2/3}_{\rm th})^2}{t^2_{\rm p}
(\omega_{fl} -\omega)^2} {\kern 2pt}\{  t^2 &-& \frac
{2t}{(\omega_{fl} -\omega)} {\kern 1pt}\sin ((\omega_{fl} -
\omega) t) \ \ \ \ \ \ \ \ \ \ \ \ \ \ \nonumber  \\ &+& \frac
2{(\omega_{fl} -\omega)^2} {\kern 1pt} [1 - \cos ((\omega_{fl} -
\omega) t)] \}. \ \ \ \ \ \ \ \label{39}
\end{eqnarray}
In our case $\omega_{fl} - \omega = (E_f-E_l) / \hbar - \omega$
where $\omega =2\pi \nu$ and $\nu$ is the frequency of incident
light.  As $\omega_{fl} \gg \omega$, one can put $\omega_{fl} -
\omega \simeq \omega_{fl}$. Besides we consider the approximation
when $t \ll t_{\rm p}= \Delta t/2 \approx 10^{-8} - 10^{-7}$ s.
Hence for the wide range of time (i.e., $\omega_{fl}^{-1} \ll t
\ll t_{\rm p}$ the inequality $\omega_{fl} t \gg 1$ is held and
expression (39) can be replaced by
\begin{equation}
|{ \cal I} (t)|^2 \simeq (\frac {\sigma_{\rm th}n^{2/3}_{\rm
th}}{\omega_{fl} {\kern 1pt}t_{\rm p}})^2 \ t^2. \label{40}
\end{equation}
The matrix element $\omega_{fl}$ in (40) can be eliminated by
substituting the absolute value of ionization potential of atoms
(or the work function of the metal) $W$, that is, $\omega_{fl}
\rightarrow W / \hbar$. If we substitute (40) into (36), we
finally get
\begin{eqnarray}
P(t)&=&| { \cal M } |^2 \ \Big( \frac {\hbar {\kern 2pt}
\sigma_{\rm th}{\kern 1pt} n^{2/3}_{\rm th}} {W t_{\rm p}}\Big)^2
\ I_{\rm p} {\kern 1pt} t^2 \nonumber
\\ &\equiv& |{\cal M}|^2 (\hbar {\cal N}/Wt_{\rm p})^2  I_{\rm
p}{\kern 1pt} t^2. \label{41}
\end{eqnarray}

In the case when the incident laser pulse one may consider as a
perturbation that is not time dependent, the interaction operator
(21) can be regarded as a constant value
\begin{eqnarray}
{\hat W}_{\rm eff}&=&-\frac{e}{m}{\kern 1pt}{\vec A}_{\rm
eff}(\vec r) {\kern 1pt}{\hat {\vec p}}       \nonumber
\\ &\equiv& -\frac{e}{m}{\cal N}{\vec A}_0 {\kern 1pt} e^{i \vec k
\vec r} {\hat {\vec p}} \label{42}
\end{eqnarray}
between the moments of cut-in and cut-off and ${\hat W}_{\rm
eff}=0$ behind  the time interval $\Delta t$ corresponding to the
duration of the laser pulse. Now having the interaction operator
(42) we directly use the Fermi golden rule and obtain the
probability of the anomalous photoelectric effect (compare with
the theory of the simple photoelectric effect, e.g. Refs. 59, 60)
\begin{equation}
P_0= \frac {2\pi}{\hbar}|{\cal M}|^2 {\cal N}^{\kern 1pt 2}  I   V
m |{\vec p}_{\rm {\kern 1pt} free}| \ \Delta t \ d \Omega;
\label{43}
\end{equation}
here $|{\cal M}|^2$ is the matrix element defined above (35),
${\cal N}= \sigma n^{2/3}$ is the number of photons absorbed by an
atom (or the metal) simultaneously, $I$ is the typical intensity
of the laser pulse, $Vm|{\vec p}_{\rm free}|$ is the density of
states ($V$ is the normalizing volume, $m$ is the electron mass
and $|{\vec p}_{\rm free}|$ is the momentum of the stripped
electron).

Thus it is easily seen that the interaction between the laser pulse and
gas atoms (or the metal) is not nonlinear. This is why the results to
be expected from this new approach would correlate with the results
predicted by the effective photon (7).

\section{\bf Discussion}

\hspace*{\parindent} Let us apply the results obtained above to
the experimental data used by Panarella [17,18] for the
verification of the effective photon. Besides other experimental
results are taken into account as well. We shall restrict our
consideration to qualitative evaluations, which note only the
general tendency towards the behaviour of the systems in question.

\subsection{\it  Laser-induced gas ionization}

\hspace*{\parindent}
   {\bf 5.1.1.} Let probability (41) describes the transition from the
stationary state of an atom to the ionized state of the same atom.
Multiplying both sides of expression (41)  by the concentration
$N_a$ of gas atoms which are found in the focal volume
investigated, one gains the formula for the concentration $N_i$ of
ionized atoms
\begin{equation}
N_i= N_a \ |{\cal M}|^2 \ \Bigl( \frac {\hbar {\cal N}_{\rm th}}
{W t_{\rm p}} \Bigr)^2 \ I_{\rm p} \ t^2.
\label{44}
\end{equation}
So, it is readily seen that
\begin{equation}
N_i \propto N_a I_{\rm p} t^2,
\label{45}
\end{equation}
that is, the concentration $N_i$ of ionized atoms is directly
proportional to the peak laser pulse intensity $I_{\rm p}$ and the
time to the second power.  Time dependence of ionization before
breakdown was analysed by Panarella [17,18] in the framework of
the same formula (45) obtained by him using the effective photon.
The experiment by Chalmeton and Papoular [40]  showed that the
evolution of free electrons knocked out of gas atoms, that is $d
\ln N_e(t)/ d t$, is only a function of time.  As the electron
density $N_e(t)=N_i(t)$, following  Refs. 17,18 we obtain from
(45) (or (44)):  $d \ln N_e(t) / dt =2/t$, in agreement with the
experiment.

  {\bf 5.1.2.} Temporal dependence of the breakdown threshold intensity
was studied by Panarella [18] with the aid of the same expression
(45). At breakdown $N_i$=const, $I_{\rm p}$ is replaced by the
threshold intensity $I_{\rm th}$ and a time interval $t$ is equal
to the breakdown time $t_{\rm b}$. Hence in this case expression
(45) gives $I_{\rm th}$=const$\times t^{-2}_{\rm b}$ or, according
to formulas (32), $E_{\rm th}\propto t^{-1}_{\rm b}$. This
expression agrees with the experiment by Pheps [52].

  {\bf 5.1.3.} The experimental results on the number of ions created
by the laser pulse as a function of the pulse intensity can also be
described in terms of the anomalous photoelectric effect. For this
purpose we should concentrate upon expression (43), which yields after
multiplying both sides by the concentration  $N_a$ of gas atoms
\begin{equation}
N_i = {\rm const} \times N_a {\cal N}^{\kern 1pt 2} I. \label{46}
\end{equation}
However before proceeding to the verification of the theory we
should call attention to the process, which is the reverse of the
photoelectric effect. The case in point is the radiation
recombination of an electron with a fixed ion, Berestetskii {\it
et al.} [61].

    The intensity $I$ of the laser pulse characterizes the density of
electromagnetic energy per unit of time, that is, one can deem
that $I$ is in inverse proportion to time. This enables the
construction of a possible model describing the occupancy of
states of ions and atoms in the presence of the strong laser
irradiation. The processes of ionization of atoms and
recombination of ions may be represented by the following kinetic
equations:
\begin{equation}
 \dot N_a = \alpha N_a - \beta N_i + D;
\label{47}
\end{equation}
\begin{equation}
 \dot N_i = \gamma N_i - \alpha N_a
\label{48}
\end{equation}
where the dot over $N_{a(i)}$ means derivation with respect to the
"time" variable $\tilde t \equiv 1/I$. Here $\alpha N_a$ and
$\beta N_i$ present the rate of ionization and restoration of
atoms of gas respectively, $\gamma N_i$ represents the rate of
recombination of ions in gas, and $D$ is the rate of irreversible
decay of the atoms (it specifies a part of electrons which leave
the gas studied). As the first approximation we can put $D=0$ and
therefore $\gamma = \beta$. Such an approximation allows the
following solution of Eqs. (47) and (48):
\begin{equation}
N_a =N_{a0} \bigl( 1-e^{-(\alpha + \beta)/I} \bigr);
\label{49}
\end{equation}
\begin{equation}
N_i = N_{a0} \Bigl( \frac{\alpha}{\beta} - \frac{\alpha}{\alpha +
2\beta} e^{-(\alpha + \beta)/I} \Bigr)
\label{50}
\end{equation}
where $N_{a0}$ is the initial concentration of atoms of gas in the
focal volume. Denote the parameter $(\alpha + \beta)$ by $I_m$,
which may correspond to an intensity supporting the balance
between ionization and recombination in the gas system studied.
Then substituting $N_a$ from the solution (49) into relation (46)
we get the resultant expression governs the total number of ions
$N_i$ as a function of the laser intensity $I$ and the number of
absorbed photons $\cal N$:
\begin{equation}
N_i = {\rm const} \times N_{a0} {\cal N}^{\kern 1pt 2} I \bigl(
1-e^{-I_m/I} \bigr). \label{51}
\end{equation}
Expression (51) correlates in outline with Panarella's [17,18]
expression which he utilized to explain the total number of ions
produced by the laser pulse (the experimental results by Agostini
{\it et al.} [39]). In fact when $I<I_m$, the exponential term can
be neglected in (51) and in a log-log plot the number of ions
versus  the pulse intensity is proportional to the number of
absorbed photons, that is
\begin{equation}
\log N_i / \log I \propto {\cal N}
\label{52}
\end{equation}
and, hence, $N_i$ against $I$ is a straight line whose slope is
$\cal N$ (see, e.g. the experimental results by Lompre {\it et
al.} [22]). When $I>I_m$, the exponent can not be neglected and,
therefore, a curve $N_i$ versus $I$ must show an inflection point
(probably at $I \simeq I_m$) in accord with the experimental
results by Agostini {\it et al.} [39].

    {\bf  5.1.4.} Expression (51) is able to explain the breakdown
intensity threshold measured as a function of pressure or gas density.
If expression (51) is written in the form
\begin{equation}
I_{\rm th}
\simeq {\rm const} \times N^{-1}_{a0} \bigl(1+ e^{-I_m/I} \bigr)
\label{53}
\end{equation}
where $I_{\rm th}$ is the breakdown threshold intensity, the
function $I_{\rm th}$ versus $N_{a0}$ indicates that $I_{\rm th}
\propto N_{a0}^{-1 + \delta}$  where the value $\delta$ satisfies
the inequalities $0< \delta < 1/2$. Such a variation of the
parameter $\delta$ rhymes satisfactory with the experimental
results by Okuda {\it et al.} [41-43] and their analysis carried
out by Panarella [17,18].

    {\bf 5.1.5.} The appearance of electrons released from atoms of gas
at high energies (more than 100 eV at the laser intensity at
$5\times 10^{14}$ W/cm$^2$, Agostini and Petite [16]) follows
immediately from the theory constructed. The two possibilities may
be realized. First of all expressions (41) and (43) allow the
kinetic energy of revealed electrons larger than $h\nu \sigma_{\rm
th}n_{\rm th}^{2/3}$ because as is evident from inequalities (12),
an electron's inerton cloud can absorb in principle more photons,
${\cal N} + \Delta {\cal N} = \sigma n^{2/3}$,  than is required
for overcoming the threshold value ${\cal N}=\sigma_{\rm th}
n_{\rm th}^{2/3}$. This is no surprise, since the anomalous
photoelectric effect is a generalization for the simple one. In
the theory of the simple photoelectric effect one can recognize
the approximations $h \nu \geq W$ and $h \nu \gg W$. The first
inequality can be related to the anomalous photoelectric effect
considered above. The second one corresponds to the Born
(adiabatic) approximation, Berestetskii {\it et  al.} [61], and in
the case of the anomalous photoelectric effect the inequality
changes merely to $({\cal N} +\Delta {\cal N})h \nu \gg W$. Notice
that this inequality is in agreement with formula (6) utilized by
the multiphoton theory to account for the energy spectrum of
electrons ejected in the ionization of atoms.

       At the same time the absorption of radiation by an accelerated
electron (called the above-threshold ionization in Ref. 16) must
not be ruled out. Actually, if a final state of a released
electron is the state of a free electron in an electromagnetic
field (so called "Volkov state" [16]), one may assume that the
electron was stripped having a very small kinetic energy. Let
initial velocity $v_0$ of the electron released from an atom be
several times less than the velocity of the electron in the atom
which we set equal to the Fermi velocity $v_{\rm F} \simeq 2\times
10^6$ m/s in Section 3. In such the case as it follows from
relation (1) and inequalities (11) the electron excites
surrounding space significantly wider than the Fermi electron and
this is why the cross section of the excited range of space around
the low speed electron should be at least ten times greater than
the magnitude of cross section evaluated in Section 3. This means
that our low speed electron will be immediately scattered by more
than ${\cal N} + 10$ photons of the laser beam and therefore its
kinetic energy may reach the value of several tens eV.

\subsection{\it Electron emission from a laser-irradiated metal }

\hspace*{\parindent}
       The investigation of the photoelectric emission from a
laser-irradiated metal performed experimentally by Panarella
[50,51,18] has shown that: \break 1) the photoelectric current
$i_e$ is linear with light intensity $I$,
\begin{equation} i_e
\propto I; \label{54}
\end{equation}
2) the maximum energy $\varepsilon_{\rm max}$ of the emitted electron is
a function of light intensity $I$,
\begin{equation}
\varepsilon_{\rm max}\propto f(I)
\label{55}
\end{equation}
and $\varepsilon_{\rm max}$ increases with $I$. The same
dependence of $i_e$ and $\varepsilon_{\rm max}$ on $I$ is
predicted by the effective photon theory [18] (note that the
multiphoton methodology predicted that $i_e$ depends on $I$  to
the power $\cal N$ and $\varepsilon_{\rm max}$ depends on $\nu$
only of the light).

Let us compare the results of the anomalous photoelectric effect theory
developed above with the experimental results by Panarella (formulas
(54) and (55)). In his experiments the light intensity $I$ changed from
$\sim 10^6$  W/cm$^2$ to $\sim 10^9$ W/cm$^2$ from experiment to
experiment. This value of $I$ is not very great and we can take into
consideration the total power transferred during one pulse.
By this is meant that the light intensity is assumed to be constant
during the pulse. Therefore the expression (43)
\begin{equation}
P_0 = {\rm const} \times (\sigma n^{2/3})^2  I
\label{56}
\end{equation}
can be used to evaluate of the electron emission from the metal.
Expression (56) was obtained utilizing the perturbation theory. In
other words, the interaction energy ${\cal W}_{\rm eff} \equiv
e{\vec A}_0 \vec p \sigma n^{2/3}/m$ that forms the perturbation
operator (42) should be smaller than the absolute value of the
work function $W$. In Panarella's experiments the value of $W$ was
about $10^{-18}$ J (i.e., approximately 6 eV). At $I=10^6 - 10^9$
W/cm$^2$ (i.e.,  $10^{24} - 10^{27}$ photons/cm$^2$ per second)
one has
\begin{equation} {\cal W}_{\rm eff}= (1.8 \times 10^{-22}
- 5.7 \times 10^{-21})\times (\sigma n^{2/3})^2 \ \  [{\rm J}].
\label{57}
\end{equation}
If we try formally to estimate an additional number of photons
$\sigma n^{2/3}-1$ which pass their energy on to the electron that
absorbed a single photon, we will find with regard for the
inequality (11):
  $$ \ \ \ \ \ \ 1<\sigma n^{2/3} \ll 1 \ \ \ \ \ \
\ \ {\rm at} \ \ \ \ n \approx 3\times 10^{15}  \ \ {\rm cm}^{-3};
\eqno(58a)
  $$
  $$ \ \ \ \ 1 < \sigma n^{2/3} < 2 \ \ \ \ \ \ \ {\rm
at}\ \ \ \ \ n \approx 10^{18}\ \ {\rm cm}^{-3}. \eqno(58b)
  $$
Substituting $\sigma n^{2/3}$ from (58$b$) in expression (57), it
is easily seen that the inequality $W\gg {\cal W}_{\rm eff}$ is
not broken, that is formula (56) could be applied to the study of
anomalous electron emission from the metal. Nonetheless,
inequalities (58$a$) are not correct while the experiment [51,18]
pointed to the presence of photoelectrons at the light intensity
$I=10^6$ W/cm$^2$ ($n \approx 3\times 10^{15}$ cm$^{-3}$). One way
around this problem is to take into account the large
concentration of electrons $n_{\rm elec}$ in a metal. Indeed, the
value of $n_{\rm elec}\sim 10^{21}$ cm$^{-3}$ and consequently the
mean distance between electrons is $n^{-1/3}_{\rm elec}\sim 1$ nm.
Bearing in mind that owing to relationship (1) the electron's
inerton cloud in the metal is characterized by amplitude $\Lambda
/\pi \simeq 17$ nm, one should supplement the parameter $\sigma$
by a correlation function $F(\Lambda, n^{1/3}_{\rm elec})$. The
function can  be  chosen in the form $$
 F= \Bigl[\frac {\Lambda}{n^{-1/3}_{\rm elec}}\Bigr]^{\gamma},
\ \ \ \ \gamma > 0. \eqno(59) $$ The function (59) corrects
inequalities (58$a$). Hence expression (56) takes the form $$
P_0={\rm const}\times (\sigma n^{2/3}F)^2 \times I \eqno(60) $$
and it can be used until $W \gg {\cal W}_{\rm eff}$. For large
$I_{\rm p}$ when ${\cal W}_{\rm eff} \sim W$, expression (60) is
also suitable, but only at the initial stage of the laser  pulse
(in this case the factor $t/t_{\rm p}$ should again be introduced
into the right hand side (60)). Note that in the case of rarefied
gases the overlapping of inerton clouds of neighboring atoms
begins for their concentration $n_{\rm atom} \geq 10^{17}$
cm$^{-3}$; here the mean distance between atoms  $n^{-1/3}_{\rm
atom} \sim 20$ nm.

        Comparing expressions (54) and  (60) we notice that they
agree: expression (60) describes the probability of the appearance
of free electrons and hence their current $i_e$ at the difference
of electric potential as a linear function of $I$.

     The behaviour of emitted electrons described by expression (55)
is consistent with the  prediction of the present theory as well.
Panarella [51] pointed out that the incident laser beam did not
heat the metal specimen. This statement is correct for the
background temperature, i.e. phonon temperature of the small
specimen. However the electron temperature should increase with
the intensity of light; it is well known phenomenon called heat
electrons (see, e.g. Refs. 62-64). The greater the light flux
intensity, the greater the kinetic energy of the heat electrons in
small metal specimens [63,64]. As a result the work function $W$
of the specimen becomes a function of the intensity of light $I$:
\ $W$ falls as $I$ increases. Thus, expression (55) should also
follow from the theory based on the inerton concept; the theory
gives the explicit form of expression (55):
 $$ \varepsilon_{\rm
 max}= {\cal N} h\nu  - W(I) \eqno(61)
 $$ where $h\nu$ is the photon
energy of incident light, ${\cal N}$ is the threshold number of
photons scattered  by the electron's inerton cloud  and $W(I)$ is
the work function depending on the intensity of light $I$.

\section{\bf Conclusion}

\hspace*{\parindent}
       The present theory of anomalous photoelectric effect has been
successfully applied to the numerous experiments where the photon
energy of incident light is essentially smaller than the
ionization potential of gas atoms and the work function of the
metal. This theory is based on submicroscopic quantum mechanics
developed in the previous papers by the author [1-3].  Note that
ideas on the microstructure of the space set forth in that
author's research are in excellent agreement with the recent
construction of a mathematical space carried out by Bounias and
Bonaly [65] and Bounias [66]. Space reveals its properties through
the engagement of the particle with it.  As a result -- a cloud of
inertons, that is, elementary excitations of the space, is created
in the surrounding of the particle and just these clouds enclosing
electrons were detected in the experiments mentioned above by a
high-intensity luminous flux.

It is obvious that clouds of inertons, which accompany electrons
were fixed also in another series of experiments carried out by a
large group of physicists, Briner {\it et al.} [67]. Their article
is entitled "Looking at Electronic Wave Functions on Metal
Surfaces" and it contains the colored spherical and elliptical
figures, which the authors called "the images of $\psi$ wave
functions of electrons". However, the wave $\psi$-function is only
a mathematical function that sets connections between parameters
of the system studied. So the wave $\psi$-function can not be
observed in principle. This means that the researchers could
register perturbations of space surrounding the electrons in the
metal, i.e., clouds of inertons accompanying moving electrons. It
is believed that mobile small deformations of space -- inertons,
which constitute a substructure of the matter waves -- promise new
interesting effects and phenomena [68,69].

    At the same time, for the description of a whole series
of phenomenological aspects of effects caused by highly intensive
laser radiation in the case when the adiabatic approximation may
be used, Panarella's effective photon theory [17,18] is also
suitable (the theory is similar to the phenomenological theory of
propagation of electromagnetic waves in nonlinear media, see, e.g.
Ref. 70). As it follows from the analysis above, the effective
photon methodology, indeed, specifies the effective photon
density, or the number of photons absorbed by the electron's
inerton cloud (see expression (15)); therefore, the methodology
allows the correct calculation of the photon energy absorbed by an
atom of gas or an electron in the metal and, as the rule, just the
value of this energy is very significant for the majority of the
problems which are researched.

    As for the nonlinear multiphoton concept, its basis should be
altered to the linear one, that is, to the anomalous photoelectric
concept developed herein.

     An important conclusion arising from the theory considered in the
present work is that the Amp\'ere's formula $\vec p - e \vec A$ is
not universal. In the general case, when the intensity of the
electromagnetic field is high, it should be  replaced by the
formula $\vec p - e {\cal N}\vec A$ where the vector potential
$\vec A$ is normalized to one photon and $\cal N$ is the quantity
of coherent photons scattering/absorbing by the electron's inerton
cloud simultaneously. In other words, for highly intensive
electromagnetic field, one should use the approximation of the
strong electron-photon coupling (see expressions (14) and (15)).

       The submicroscopic approach is not only advantageous in the
study of matter under strong laser irradiation. The approach
provides a means of more sophisticated analysis of the nature of
matter waves and the nature of light. Thereby such an analysis is
able to originate radically new view-points on the structure of
real space, the notions of particle and field and their
interaction.

\vspace{4mm}

{\bf Acknowledgement}

\vspace{2mm}

       I am very thankful to Dr. E. Panarella who provided me
with his reviews, which were used as a basis for the paper
presented herein and I would like to thank to Prof. M. Bounias for
the fruitful discussion concerning the background of the developed
concept. And I am very thankful to Mrs. Gwendolin Wagner who paid
the page charge for the publication of the present work.

\vspace{8mm}

\ {\bf References}

\begin{enumerate}

\item Krasnoholovets, V.  and Ivanovsky, D. -- Motion  of a
particle and the vacuum, {\it Phys. Essays} {\bf 6}, 554-563
(1993) (also arXiv.org e-print archive,
http://arXiv.org/abs/quant-ph/9910023).

\item Krasnoholovets, V. -- Motion of a relativistic particle
and the vacuum, {\it Phys. Essays} {\bf 10}, 407-416 (1997) (also
quant-ph/9903077).

\item V. Krasnoholovets, -- On the nature of spin, inertia and
gravity of a moving canonical particle, {\it Ind. J. Theor. Phys.}
{\bf 48}, 97-132 (2000)  (also quant-ph/0007027).

\item De Broglie, L. --  Interpretation  of quantum mechanics
by the double solution theory, {\it Ann. de la Fond. L. de
Broglie} {\bf 12}, 399-421 (1987).

\item Bohm, D. -- A suggested interpretation of the quantum
theory, in terms of "hidden" variables. I, {\it Phys. Rev.} {\bf
85}, 166-179 (1952); A suggested interpretation of the quantum
theory, in terms of "hidden" variables. II, {\it ibid.} {\bf 85},
180-193 (1952).

\item  Rado, S. -- {\it Aethero-kinematics}, CD-ROM (1994), Library
of Congress  Catalog Card,  \# TXu 628-060 (also
http://www.aethero-kinematics.com).

\item Aspden, H. --  {\it Aetherth science papers}, Subberton
Publications,  P. O. Box 35, Southampton SO16 7RB, England (1996).

\item Kohler, C. -- Point particles in 2+1 dimensional gravity
as defects in solid continua, {\it Class. Quant. Gravity} {\bf
12}, L11-L15 (1995).

\item  Vegt, J. W. -- A  particle-free model of matter based
on electromagnetic self-confinement (III), {\it Ann. de la Fond.
L. de Broglie} {\bf 21}, 481-506 (1996).

\item Winterberg, F. -- {\it The Planck aether hypothesis. An
attempt for a finistic theory of elementary particles}, Verlag
relativistischer Interpretationen -- VRI, Karlsbad (2000).

\item Rothwarf, A. --  An aether model of the universe, {\it Phys.
Essays} {\bf 11}, 444-466 (1998).

\item Berezinskii, V. S. -- Unified gauge theories and
unstable proton, {\it Priroda} (Moscow), no. 11 (831), 24-38
(1984) (in Russian).

\item Meyerand, R. G., and Haught, A. F. -- Gas breakdown at
optical frequencies, {\it Phys. Rev. Lett.} {\bf 11}, 401-403
(1963).

\item Voronov, G. S.,  and Delone, N. B. --  Ionization of
xenon atom by electric field of ruby laser radiation, {\it JETP
Lett.} {\bf 1}, no. 2, 42-45 (1965) (in Russian).

\item Smith, D. C., and Haught, A. F. -- Energy-loss
processes in optical-frequency gas breakdown, {\it Phys. Rev.
Lett.} {\bf 16}, 1085-1088 (1966).

\item Agostini, P., and Petite, G., -- Photoelectric effect
under strong irradiation, {\it Contemp. Phys.} {\bf 29}, 57-77
(1988).

\item Panarella, E. --  Theory of laser-induced gas
ionization, {\it Found. Phys.} {\bf 4}, 227-259 (1974).

\item Panarella, E. -- Effective photon hypothesis vs.
quantum potential theory: theoretical predictions and experimental
verification, in: {\it Quantum uncertainties. Recent and future
experiments and interpretations}. NATO ASI.  Series B 162,
Physics, eds.:  Honig, W. M., Kraft, D. W. and Panarella, E.,
Plenum Press, New York (1986),  237-269.

\item Fermi, E. -- {\it Notes on quantum mechanics}, Mir, Moscow
(1965), p. 211 (Russian translation).

\item Keldysh, L. V. -- Ionization in the field of a strong
electromagnetic wave, {\it JETP} {\bf 47}, 1945-1957 (1964) (in
Russian).

\item Reiss, H. R. -- Semiclassical electrodynamics of bound
systems in intense fields, {\it Phys. Rev. A} {\bf 1}, 803-818
(1970).

\item Lompre, L. A., Mainfray, G., Manus, C., and Thebault,
J. -- Multiphoton ionization of rare gases by a tunable-wavelength
30-psec laser pulse as 1.06 $\mu$m, {\it Phys. Rev. A} {\bf 15},
1604-1612 (1977).

\item Martin, E. A. and Mandel, L. -- Electron energy
spectrum in laser-induced multiphoton ionization of atoms, {\it
Appl. Opt.} {\bf 15}, 2378-2380 (1976).

\item Boreham, B. W., and Hora, H. -- Debye-length
discrimination of nonlinear laser forces acting on electrons in
tenuous plasmas, {\it Phys. Rev. Lett.} {\bf 42}, 776-779 (1979).

\item Petite, G., Fabre, F., Agostini, P., Grance, M., and
Aymar, M., -- Nonresonant multiphoton ionization of cesium in
strong fields: angular distributions and above-threshold
ionization, {\it Phys. Rev. A} {\bf 29}, 2677-2689 (1984).

\item Kruit, P., Kimman, J., and Van der Wiel, M. J. --
Absorption of additional photons in the multiphoton ionization
continuum of xenon at 1064, 532 and 440 nm,  {\it J. Phys. B} {\bf
14}, L597-602 (1981).

\item Fabre, F.,  Petite, G., Agostini, P., and Clement, M.
-- Multiphoton above-threshold ionization of xenon at 0.53 and
1.06 $\mu$m, {\it J. Phys. B} {\bf 15}, 1353-1369 (1982).

\item  Agostini, P., Kupersztych, J., Lompre, L. A., Petite,
G., and Yergeau, F. -- Direct evidence of ponderomotive effect via
laser pulse duration in above-threshold ionization, {\it Phys.
Rev. A} {\bf 36}, 4111-4114 (1987).

\item Farkas, G.  in: {\it Photons and continuum states of atoms
and molecules}, eds.: N. K. Rahman, C. Guidotti and M. Allegrini,
Springer-Verlag, Berlin (1987), p. 36.

\item Fedorov, M. V. -- {\it An electron in strong light field},
Nauka, Moscow, (1991) (in Russian).

\item Manfray, G., and  Manus, C. -- Multiphoton ionization of
atoms, {\it Rep. Prog. Phys.} {\bf 54}, 1333-1372 (1991).

\item Mittleman, M. H. -- {\it Introduction to the theory of
laser-atom interactions}, Plenum, New York, (1993).

\item Delone, N. B.,  and Krainov, V. P. -- {\it Multiphoton
processes in atoms}, Springer, Heidelberg (1994).

\item Delone, N. B., and Krainov, V. P. -- Stabilization of
an atom by the field of laser radiation, {\it Usp. Fiz. Nauk} {\bf
165},  1295-1321 (1995) (in Russian).

\item Avetissian, H. K., Markossian, A. G., and Mkrtchian, G.
F. -- Relativistic theory of the above-threshold multiphoton
ionization of hydrogen-like atoms in the ultrastrong laser fields,
quant-ph/9911070.

\item Protopapas, M., Keitel, C. H., and Knight, P. L. --
Atomic physics with super-high intensity lasers, {\it Rep. Prog.
Phys.} {\bf 60}, 389 (1997).

\item Salamin, Y. I. -- Strong-field multiphoton ionization
of hydrogen: Nondipolar asymmetry and ponderomotive scattering,
{\it Phys. Rev. A} {\bf 56}, 4910-4917 (1997).

\item   Avetissian, H. K.,  Markossian, A. G., Mkrtchian, G.
F., and  Movsissian, S. V. -- Generalized eikonal wave function of
an electron in stimulated bremsstrahlung in the field of a strong
electromagnetic wave, {\it Phys. Rev. A} {\bf 56}, 4905-4909
(1997).

\item Agostini, P.,  Barjot, G., Mainfray, G., Manus, C., and
Thebault, J. -- Multiphoton ionization of rare gases at 1.06
$\mu$m and 0.53 $\mu$m,  {\it IEEE J. Quant. Electr.} {\bf QE-6},
782-788 (1970).

\item Chalmeton, V., and Papoular, R. -- Emission of light by
a gas under the efeect of an intense laser radiation,  {\it Compt.
Rend.} {\bf 264B}, 213-216 (1967).

\item Okuda, T., Kishi, K., and Savada, K. -- Two-photon
ionization process in optical breakdown of cesium vapor, {\it
Appl. Phys. Lett.} {\bf 15}, 181-183 (1969).

\item Kishi, K., Sawada, K., Okuda, T., and Matsuoka, Y. --
Two-photon ionization of cesium and sodium vapors, {\it J. Phys.
Soc. Jap.} {\bf 29}, 1053-1061 (1970).

\item Kishi, K., and Okuda, T. -- Two-photon ionization of
alkali metal vapors by ruby laser,  {\it J. Phys. Soc. Japan} {\bf
31}, 1289 (1971).

\item  Zel'dovich, Ya. B., and Raizer, Yu. P. -- Cascade
ionization of a gas by a light pulse, {\it JETP} {\bf 47},
1150-1161 (1964).

\item Allen, A. D. -- A testable Noyes-like interpretation of
Panarella's effective-photon theory, {\it Found. Phys.} {\bf 7},
609-615 (1977).

\item Dewdney, C., Garuccio, A., Kyprianidis, A., and Vigier,
J. P. -- The anomalous photoelectric effect: quantum potential
theory versus effective photon hypothesis, {\it Phys. Lett.} {\bf
105A}, 15-18 (1984).

\item Dewdney, C.,  Kyprianidis, A., Vigier, J. P., and
Dubois, A. -- Causal stochastic prediction of the nonlinear
photoelectric effects in coherent intersecting laser beams, {\it
Lett. Nuovo Cim.} {\bf 41}, 177-185 (1984).

\item De Brito, A. L.,  and Jabs, A. -- Line broadening by
focusing, {\it Can. J. Phys.} {\bf 62}, 661-668 (1984).

\item De Brito, A. L. -- Gas ionization by focused laser
beams, {\it Can. J. Phys.} {\bf 62}, 1010-1013 (1984).

\item Panarella, E. -- Experimental test of multiphoton
theory, {\it Lett. Nuovo Cim.} {\bf 3}, Ser.2, 417-423 (1972).

\item Panarella, E. -- Spectral purity of high-intensity
laser beams,  {\it Phys. Rev. A} {\bf 16}, 672-680 (1977).

\item Pheps, A. V. -- Theory of growth of ionization during
laser breakdown, in: {\it Physics of quantum electronics}, eds. P.
L. Kelley, B. Lax and P. E. Tannelwald, McGraw-Hill Book Company,
New York (1966), 538-547.

\item Raychaudhuri, P. -- Effective photon hypothesis and the
structure of the photon, {\it Phys. Essays} {\bf 2}, pp. 339-345
(1989).

\item Berestetskii, V. B., Lifshitz, E. M.,  and Pitaevskii,
L. P. -- {\it Quantum electrodynamics}, Nauka, Moscow (1980), p.
28 (in Russian).

\item Davydov, A. S.-- {\it The theory of solids}, Nauka, Moscow
 (1976), p. 350 (in Russian).

\item Davydov, A. S. -- {\it Quantum mechanics}, Nauka, Moscow
(1973),  p. 374 (in Russian).

\item Ter Haar, D. -- {\it Elements of Hamiltonian mechanics},
Nauka, Moscow (1973), p. 374 (Russian translation).

\item See Ref. 54, p. 231.

\item Blokhintsev, D. I. -- {\it Principles of quantum mechanics},
Nauka, Moscow (1976), p. 407 (in Russian).

\item See Ref. 56, p. 472.

\item See Ref. 54, p. 242.

\item Anisimov, S. I., Imos, Ya. A., Romanov, G. S., and
Khodyko, Yu. V. -- {\it Action of high-intensity radiation on
metals}, Nauka, Moscow (1970) (in Russian).

\item Tomchuk, P. M. -- Electron emission from island metal
films under the action of laser infrared radiation (theory), {\it
Izvestia Acad. Sci. UdSSR}, Ser. Phys. {\bf 52}, 1434-1440
(1988)(in Russian).

\item Belotsky, E. D., and  Tomchuk, P. M. -- Electron-photon
interaction and hot electrons in smal metal islands,  {\it Surface
Sci.} {\bf 239}, 143-155 (1990).

\item Bounias, M.  and Bonaly, A. -- On mathematical links
between physical existence, observebility and information: towards
a "theorem of something",  {\it Ultra Scientist of Phys. Sci.}
{\bf 6}, 251-259 (1994); \  Timeless space is provided by empty
set, {\it ibid.} {\bf 8}, 66-71 (1996); \ On metric and scaling:
physical co-ordinates in topological spaces, {\it Ind. J. Theor.
Phys.} {\bf 44}, 303-321 (1996); \ Some theorems on the empty set
as necessary and sufficient for the primary topological axioms of
physical existence,  {\it Phys. Essays} {\bf 10}, 633-643 (1997).

\item Bounias, M. --  The theory of something: a theorem
supporting the conditions for existence of a physical universe,
from the empty set to the biological self, {\it Int. J. Anticip.
Syst.} {\bf 5-6}, 1-14 (2000).

\item  Briner, G.,  Hofmann, Ph. Doering, M., Rust, H. P.,
Bradshaw, A. M.,  Petersen, L.,  Sprunger, Ph.,  Laegsgaard, E.,
Besenbacher, F. and  Plummer, E. W. -- Looking at electronic wave
functions on metal surfaces,  {\it Europhys. News} {\bf 28},
148-152 (1997).

\item Krasnoholovets, V., and Byckov, V. -- Real inertons
against hypothetical gravitons. Experimental proof of the
existence of inertons, {\it Ind. J. Theor. Phys.} {\bf 48}, 1-23
(2000).

\item Krasnoholovets, V., and Lev, B. -- Systems of particles
with interaction and the cluster formation in condensed matter,
{\it Conden. Matt. Phys.}, in press.

\item Vinogradova, M. V., Rudenko, O. V.,  and Sukhorukov, A.
P. --  {\it The theory of waves}, Nauka, Moscow (1979) (in
Russian).
\end{enumerate}

\end{document}